\documentclass[twocolumn,pre,amsmath,amssymb]{revtex4}

\usepackage{graphicx}
\usepackage{latexsym}
\usepackage{amsmath}

\begin{document}

\title{Models of wealth distributions: a perspective}

\author{Abhijit Kar Gupta}

\affiliation{Physics Department, Panskura Banamali College\\
Panskura, East Midnapore, W.B., Pin-Code: 721 152\\
India\\
e-mail:~abhijit$_{-}$kargupta@rediffmail.com, abhijit.kargupta@saha.ac.in}

\maketitle

\date{\today}

\section {Abstract}\label{akg:sec:abs}

A class of conserved models of wealth distributions are studied where wealth (or money) 
is assumed to be exchanged between a pair of agents in a population just like the 
elastically colliding molecules of a gas exchanging energy. All sorts of distributions 
from exponential (Boltzmann-Gibbs) to something like Gamma distributions and to that of Pareto's law 
(power law) are obtained out of such models with simple algorithmic exchange processes. 
Numerical investigations, analysis through transition matrix and a mean field approach 
are employed to understand the generative mechanisms. A general scenario is examined wherefrom 
a power law and other distributions can emerge.

\section{Introduction}\label{akg:sec:intro}

Wealth is accumulated in various forms and factors. The continual exchange of wealth 
(a value assigned) among the agents in an economy gives rise to interesting and many often 
universal statistical distributions of individual wealth. Here the word `wealth' is used 
in a general sense for the purpose and the spirit of the review (inspite of 
separate meanings attached to the terms `money', `wealth' and `income').
Econophysics of wealth distributions \cite{AKG:eco} is an emerging area where mainly the 
ideas and techniques of statistical physics are used in interpreting real economic data of 
wealth (available mostly in terms of income) of all kinds of people or other entities ({\em e.g.}, 
companies) for that matter, pertaining to different societies and nations. 
Literature in this area is growing steadily (see an excellent website \cite{AKG:ecoforum}). 
The prevalence of income data and apparent interconnections of many socio-economic problems 
with physics have inspired a wide variety of statistical models, data analysis and other 
works in econophysics \cite{AKG:stanley}, sociophysics and other emerging areas \cite{AKG:stauffer} 
in the last one decade or more (see an enjoyable article by Stauffer \cite{AKG:outside}). 

Simple approach of {\it agent based models}  
have been able to bring out all kinds of wealth distributions that open up a whole new way of 
understanding and interpreting empirical data. 
One of the most important and controversial issues has been to understand the emergence of 
{\em Pareto's law}: 
\begin{equation}\label{akg:eqn:pareto}
P(w) \propto w^{-\alpha},
\end{equation}

\noindent
where $w\ge w_0$, $w_0$ being some value of wealth beyond which the power law 
is observed (usually towards the tail of the distributions). Pareto's law has been observed 
in income distributions among the people of almost all kinds of social systems across the 
world in a robust way. This phenomenon is now known for more than a century and has been 
discussed at a great length in innumerable works in economics, econophysics, sociophysics and 
physics dealing with power law distributions.
In many places, while mentioning 
{\it Pareto's law}, the power law is often written 
in the form: $P(w) \propto w^{-(1+\nu)}$, where $\nu$ is referred to as `Pareto index'. 
This index is usually found between 1 and 2 from empirical data fitting.
Power laws in distributions appear in many other cases \cite{AKG:lognormal, AKG:newman, AKG:reed} like that of 
computer file sizes, the growth of sizes of business firms and cities {\em etc}. 
Distributions are often referred to as `heavy tailed' or `fat tailed' 
distributions \cite{AKG:mandelbrot}. 
Smaller the value of $\alpha$, fatter the tail of the distribution as it may easily be 
understood (the distribution is more spread out). 

Some early attempts \cite{AKG:early} have been made to understand the income distributions 
which follow Pareto's law at the tail of the distributions. 
Some of them are stochastic
logistic equations or some related generalized versions of that which have been able to 
generate power laws. However, the absence of direct interactions of one agent with any other 
often carries less significance in terms of interpreting real data.   

Some part of this review is centered around the concept of emergence of Pareto's law 
in the wealth distributions, especially in the context of the models that are discussed here.
However, a word of caution is in order. 
In the current literature and as well as in the historical occurrences, the power law distribution has often
been disputed with a closely related lognormal distribution \cite{AKG:lognormal}. 
It is often not easy to distinguish between the two. Thus a brief discussion is made here on
this issue. Let us consider the probability density function of a lognormal distribution:
\begin{equation}\label{akg:eqn:lognormal-1}
p(w)={1\over \sqrt{2\pi}\sigma w}\exp[-(\ln w -\overline w)^2/{2\sigma^2}],
\end{equation}

\noindent
The logarithm of the above can be written as:
\begin{equation}\label{akg:eqn:lognormal-2}
\ln p(w)=-\ln w-\ln\sqrt{2\pi}\sigma-{(\ln w-\overline w)^2\over 2\sigma^2}. 
\end{equation}

\noindent
If now the variance $\sigma^2$ in the lognormal distribution is large enough, the last term 
on the right hand side can be very small so that the distribution may appear linear on a 
log-log plot. Thus the cause of concern remains, particularly when one deals with real data.  

In the literature, sometimes one calculates a {\it cumulative distribution} function 
(to show the power law in a more convincing way) instead of plotting 
ordinary distribution from simple histogram (probability density function). 
The cumulative probability distribution function $P(\ge w)$ is such that the argument has a 
value greater than or equal to $w$:
\begin{equation}\label{akg:eqn:discum-1}
P(\ge w) = \int_w^{\infty} P(w^{\prime})dw^{\prime}.
\end{equation}

\noindent
If the distribution of data follows a power law $P(w)=Cw^{-\alpha}$, then 
\begin{equation}\label{akg:eqn:discum-2}
P(\ge w) = C\int_w^{\infty} {w^{\prime}}^{-\alpha}dw^{\prime}={C\over \alpha-1}w^{-(\alpha-1)}.
\end{equation}

\noindent
When the ordinary distribution (found from just histogram and binning) is a 
power law, the cumulative distribution thus also follows a power law with the exponent 1 less: 
$\alpha-1$, which can be seen from a log-log plot of data. An extensive discussion on power laws 
and related matters can be found in \cite{AKG:newman}.  

Besides power laws, a wide variety of wealth distributions from exponential to 
something like Gamma distributions are all reported in recent literature in econophysics.
Exchange of wealth is considered to be a primary mechanism behind all such distributions.
In a class of wealth exchange models \cite{AKG:chak1, AKG:chak2, AKG:yako1, AKG:yako2} 
that follow, the economic activities among agents have been assumed 
to be analogous to random elastic collisions among molecules as considered in kinetic gas 
theory in statistical physics. 
Analogy is drawn between wealth ($w$) and Energy ($E$), where 
the average individual wealth ($\overline w$) at equilibrium is equivalent to temperature ($T$). 
Wealth ($w$) is assumed to be exchanged between two randomly selected economic agents 
like the exchange of energy between a pair of molecules in kinetic gas theory. 
The interaction is such that one agent wins and the other loses the 
same amount so that the sum of their wealth remains constant before and after an
interaction (trading): $w_i(t+1)+w_j(t+1)=w_i(t)+w_j(t)$; each trading increases time $t$ by one unit. 
Therefore, it is basically a process of zero sum exchange between a pair 
of agents; amount won by one agent is equal to the amount lost by another.
This way wealth is assumed to be redistributed among a fixed number 
of agents ($N$) and the local conservation ensures the total wealth ($W = \sum w_i$) of all the 
agents to remain conserved. 

Random exchange of wealth between a randomly selected pair of agents may be viewed as 
a {\it gambling process} (with zero sum exchange) which leads to Boltzmann-Gibbs type 
exponential distribution in individual wealth ($P(w) \propto \exp(-w/{\overline w}$). 
However, a little variation in the mode of wealth exchange can lead to a distribution 
distinctly different from exponential. A number of agent based conserved models  
\cite{AKG:chak2,AKG:saving1,AKG:saving2,AKG:models,AKG:chak3,AKG:akg2,AKG:prefer,AKG:sita}, invoked in recent times, are essentially 
variants of a gambling process. A wide variety of distributions evolve out of these models. 
There has been a renewed interest in such two-agent exchange models
in the present scenario while dealing with various problems in social systems involving 
complex interactions. 
A good insight can be drawn by looking at the 
$2\times 2$ transition matrices associated with the process of wealth 
exchange \cite{AKG:akg1}. 

In this review, the aim would be to arrive at some understanding of 
how wealth exchange processes in a simple working way may lead to a variety of 
distributions within the framework of the conserved models
A fixed number of $N$ agents in a system are allowed to 
interact (trade) stochastically and thus wealth is exchanged between them. 
The basic steps of such a wealth exchange model are as follows: 
\begin{equation} \label{akg:eqn:basic} 
w_i(t+1)=w_i(t)+\Delta w, 
\end{equation}
\begin{equation*}
w_j(t+1)=w_j(t)-\Delta w,
\end{equation*}

\noindent 
where $w_i(t)$ and $w_j(t)$ are wealths of $i$-th and $j$-th agents at time $t$ and 
$w_i(t+1)$ and $w_j(t+1)$ are that at the next time step ($t+1$).
The amount $\Delta w$ (to be won or to be lost by an agent) is determined by the nature of 
interaction. 
If the agents are allowed to interact for a long enough time, a steady state equilibrium 
distribution for individual wealth is achieved. 
The equilibrium distribution does not depend on the initial configuration (initial 
distribution of wealth among the agents). 
A single interaction between a randomly chosen pair of 
agents is referred here as one `time step'. In some simulations, $N$ such interactions 
are considered as one time step. This, however, does not matter as long as the system is 
evolved through enough time steps to come to a steady state and then data is collected for 
making equilibrium probability distributions. 
For all the numerical results presented here, data have been produced following the available models, 
conjectures and conclusions. Systems of $N=1000$ agents have been considered in each case. In each 
numerical investigation, the system is allowed to equilibrate for a sufficient time 
that ranges between $10^5$ to $10^8$ time steps. 
Configuration averaging 
has been done over $10^3$ to $10^5$ initial configurations in most cases. 
The average wealth (averaged over the agents) is kept fixed at $\overline w=1$ 
(by taking total wealth, $W=N$) for all 
the cases. The wealth distributions, that are dealt here in this review, 
are ordinary distributions (probability density function) and not the cumulative ones.

\section{Pure gambling}\label{akg:sec:pure}

\begin{figure}[htb]
\includegraphics[width=.4\textwidth]{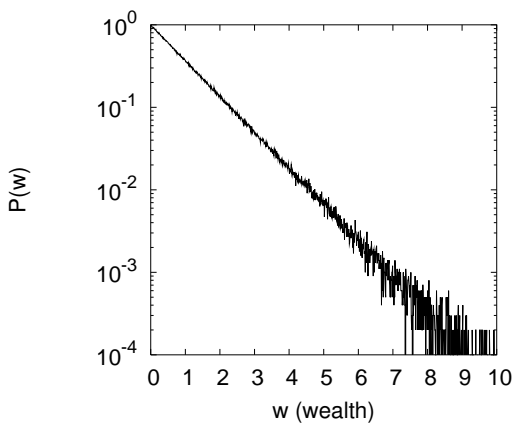}
\caption{Distribution of wealth for the case of Pure Gambling: the linearity in the 
semi-log plot indicates exponential distribution.\label{akg:fig:dist_expo}} 
\end{figure}

In a pure gambling process (usual kinetic gas theory), entire sum of wealths of two agents is up 
for gambling. Some random fraction of this sum is shared by one agent and the rest goes to 
the other. The randomness or stochasticity is introduced into the model through a 
parameter $\epsilon$ which is a random number drawn from a uniform distribution in [0, 1]. 
(Note that $\epsilon$ is independent of a pair of agents {\em i.e.}, a pair of 
agents is not likely to share the same fraction of aggregate wealth the same way when they
interact repeatedly). 
The interaction can be seen through:
\begin{equation}\label{akg:eqn:gamble}
w_i(t+1)=\epsilon[w_i(t)+w_j(t)], 
\end{equation}
\begin{equation*}
w_j(t+1)=(1-\epsilon)[w_i(t)+w_j(t)],
\end{equation*}

\noindent
where the pair of agents (indicated by $i$ and $j$) are chosen randomly. The amount of wealth that is 
exchanged is $\Delta w = \epsilon[w_i(t) + w_j(t)] - w_i(t)$.
The individual wealth distribution ($P(w)$ vs. $w$) at equilibrium emerges out to be 
Boltzmann-Gibbs distribution like exponential. 
Exponential distribution of personal income has in fact been shown to appear in real data 
\cite{AKG:yako1, AKG:yako2}. In the kinetic theory model, the exponential distribution is found by 
standard formulation of master equation or by entropy maximization method, the latter has been
discussed later in brief in section \ref{akg:sec:ineq}.
A normalized exponential distribution obtained numerically out of this pure gambling process is 
shown in Fig.~\ref{akg:fig:dist_expo} in semi logarithmic plot.
The high end of the distribution appears noisy due to sampling of data. The successive 
bins on the right hand side of the graph contain less and less number of samples in them 
so the fractional counts in them are seen to fluctuate more (finite size effect). One way to get rid of this 
sampling error in a great extent is by way of taking logarithmic binning \cite{AKG:newman}. 
Here it is not important to do so as the idea is to show the nature of the curve only. 
(In the case of power law distribution, an even better way to demonstrate and extract the 
power law exponent is to plot the cumulative distribution as discussed already.)  

If one intends to take time average of wealth of a single agent over a sufficient time, it
comes out to be equal for all the agents. Therefore, the distribution of 
individual {\it time averaged wealth turns out to be a delta function} which is checked 
from numerical data. This is because the fluctuation of wealth of any agent over time 
is statistically no different from that of any other. The same is true in case of the
distribution of wealth of a single agent over time. However, when the average of wealth of 
any agent is 
calculated over a short time period, the delta function broadens and the right end part of which 
decays exponentially. The distribution of individual wealth at a certain time turns out to 
be purely exponential as mentioned earlier. This may be thought of as a `snap shot' 
distribution.  

\section{Uniform saving propensity}\label{akg:sec:fixlam}

Instead of random sharing of their aggregate wealth during each interaction, if the agents 
decide to save (keep aside) a uniform (and fixed) fraction ($\lambda$) of their current 
individual wealth, then the wealth exchange equations look like the following:
\begin{equation}\label{akg:eqn:eqsave}
w_i(t+1)=\lambda w_i(t)+\epsilon(1-\lambda)[w_i(t)+w_j(t)], 
\end{equation}
\begin{equation*}
w_j(t+1)=\lambda w_j(t)+(1-\epsilon)(1-\lambda)[w_i(t)+w_j(t)], 
\end{equation*}

\noindent
where the amount of wealth that is exchanged is 
$\Delta w = (\epsilon-1)(1-\lambda)[w_i(t)+w_j(t)]$.
The concept of saving as introduced by Chakrabarti and group \cite{AKG:chak2} in an otherwise 
gambling kind of interactions brings out distinctly different distributions. 
A number of numerical works followed \cite{AKG:gamma, AKG:gamma-support, AKG:sudha}
in order to understand the emerging distributions to some extent. 
Saving induces accumulation of wealth. Therefore, it is expected that the probability of finding 
agents with zero wealth may be zero unlike in the previous case of pure gambling 
where due to the unguarded nature of exchange many agents are likely to go nearly bankrupt!
(It is to be noted that for an exponential distribution, the peak is at zero.)
In this case the most probable value of the distribution (peak) is somewhere else than at 
zero (the distribution is right skewed). 
The right end, however, decays exponentially for large values of $w$. 
It has been claimed through heuristic arguments (based on numerical results) that the distribution 
is a close approximate form of the Gamma distribution \cite{AKG:gamma}: 
\begin{equation}\label{akg:eqn:gamma-1}
P(w) = {n^n\over \Gamma(n)}w^{n-1}e^{-nw},
\end{equation}

\noindent
where the Gamma function $\Gamma(n)$ and the index $n$ are understood to be related to the 
saving propensity parameter $\lambda$ through the following relation:
\begin{equation}\label{akg:eqn:gamma-2}
n = 1 + {3\lambda\over 1-\lambda}.
\end{equation}

\noindent
The emergence of probable Gamma distribution is also subsequently supported through numerical 
results in \cite{AKG:gamma-support}. However, it has later been 
shown in \cite{AKG:moments} by considering moments' equation that moments up to third order agree 
with that obtained from the above form of distribution subject to the condition stated in
eqn.~(\ref{akg:eqn:gamma-2}). Discrepancies start showing only from 4th 
order onwards. Therefore, the actual form of distribution still remains an open question. 

\begin{figure}[htb]
\includegraphics[width=.4\textwidth]{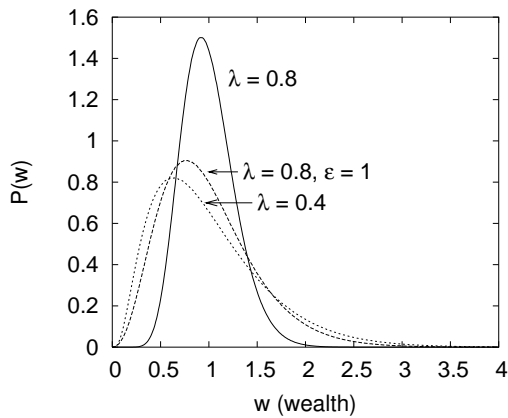}
\caption{Wealth distribution for the model of uniform and fixed saving propensity. 
Two distributions are shown with $\lambda=0.4$ and $\lambda=0.8$ where the stochasticity 
parameter $\epsilon$ is drawn randomly and uniformly in [0, 1]. Another distribution is
plotted with $\lambda = 0.8$ but with fixed value of the stochasticity parameter, 
$\epsilon = 1$.
\label{akg:fig:dist_fixlam}}
\end{figure}

In Fig.~\ref{akg:fig:dist_fixlam}, two distributions are shown for two different values of 
saving propensity factor: $\lambda=0.4$ and $\lambda=0.8$. Smaller the value of $\lambda$, 
lesser the amount one is able to save. This in turn means more wealth is available in 
the market for gambling. In the limit of zero saving ($\lambda=0$) the model reduces to that
of pure gambling. In the opposite extent of large saving, only a small amount of wealth  
is up for gambling. Then the exchange of wealth will not be able to drastically change the 
amount of individual wealth. This means the width of distribution of individual wealth 
will be narrow. In the limit of $\lambda=1$, all the agents save all of their wealth and thus
the distribution never evolves.
The concept of `saving' here is of course a little different from that in real life where 
people do save some amount to be kept in a bank or so and 
the investment (or business or gambling) is done generally not with the entire 
amount (or a fraction) of wealth that one holds at a time.

Stochastic evolution of individual wealth is also examined without the inclusion of 
stochastic parameter $\epsilon$. The stochasticity seems to be automatically 
introduced anyhow through the random selection of a pair of agents (and the random choice
of the winner or loser as well) at each time. Therefore, it is 
interesting to see how the distributions evolve with a fixed value of $\epsilon$. 
As an example, the equations in (\ref{akg:eqn:eqsave}) reduce to the following with
$\epsilon = 1$:
\begin{equation} \label{akg:eqn:eqsave_reduce}
w_i(t+1)=w_i(t)+(1-\lambda)w_j(t),
\end{equation}
\begin{equation*}
w_j(t+1)=\lambda w_j(t).
\end{equation*}

\noindent
The above equations indicate that the randomly selected agent $j$ 
keeps (saves) an amount $\lambda w_j(t)$ which is proportional to the wealth he currently has 
and transfers the rest to the other agent $i$. This is indeed a stochastic process and 
is able to produce Gamma type distributions in wealth as observed. However, a distribution with 
random $\epsilon$ and that with a fixed $\epsilon$ are different. Numerically, it 
has been observed, the distribution with $\lambda = 0.8$ and with 
$\epsilon = 1$ is very close to that with $\lambda = 0.5$ and with random $\epsilon$. 
In Fig.~\ref{akg:fig:dist_fixlam} the distribution with fixed $\lambda = 0.8$ and 
fixed $\epsilon = 1$ is plotted along with other two distributions with random $\epsilon$. 
It should also be noted that while with fixed $\epsilon$, one does not get Gamma type 
distributions 
for all values of $\lambda$; especially for low values of $\lambda$ the distributions 
become close to exponential as observed. This is not clearly understood though.

It has recently been brought to notice in \cite{AKG:lux} that a very similar kind of agent 
based model was proposed by Angle \cite{AKG:angle} (see other references cited in \cite{AKG:lux}) in 
sociological journals quite some years ago. The pair of equations in Angle's model are as
follows: 
\begin{equation}\label{akg:eqn:angle}
w_i(t+1)=w_i(t)+D_t\omega w_j(t)-(1-D_t)\omega w_i(t), 
\end{equation}
\begin{equation*}
w_j(t+1)=w_j(t)+(1-D_t)\omega w_i(t)-D_t\omega w_j(t), 
\end{equation*}

\noindent 
where $\omega$ is a fixed fraction and the winner is decided through a random 
toss $D_t$ which takes a value either 0 or 1. 
Now, the above can be seen as the more formal way of writing the pair of equations
(\ref{akg:eqn:eqsave_reduce}) which can be arrived at by choosing $D_t=1$ and  
identifying $\omega=(1-\lambda)$.

It can in general be said, within the framework of this kind of (conserved) models, 
different ways of incorporating wealth exchange processes may lead to drastically different
distributions. If the gamble is played 
in a {\it biased way}, then this may lead to a distinctly different situation than the 
case when it is played in a normal unbiased manner. 
Since in this class 
of models negative wealth or debt is not allowed, it is desirable that in each wealth 
exchange, the maximum that any agent may invest is the amount that he has at that time. 
Suppose, the norm is set for an `equal amount invest' where the amount to be deposited 
by an agent for gambling is decided by the amount the poorer agent can afford and consequently
the same amount is agreed to be deposited by the richer agent. Let us suppose $w_i > w_j$. 
Now, the poorer agent ($j$) may invest a certain fraction of his wealth, an 
amount $\lambda w_j$ and the rest $(1-\lambda)w_j$ is saved by him. Then the total 
amount $2\lambda w_j$ is up for gambling and as usual a fraction of 
this, $2\epsilon\lambda w_j$ may be shared by the richer agent $i$ where the 
rest $(1-\epsilon)\lambda w_j$ goes to the poorer agent $j$. This may appear fine, however, 
this leads to `rich gets richer and poor gets poorer' way. 
The richer agent draws more and more wealth in his favour in the successive encounters 
and the poorer agents are only able to save less and less and finally there is a 
condensation of wealth at the hands of the richest person. 
This is more apparent when one considers an agent with $\lambda = 1$ where it can be 
easily checked that the richer agent automatically saves an amount equal to the difference 
of their wealth ($w_i-w_j$) and the poorer agent ends up saving zero amount. Eventually, 
poorer agents get extinct. This is `minimum exchange model' \cite{AKG:sita}. 

\section{Distributed saving propensity}\label{akg:sec:ranlam}

The distributions emerge out to be dramatically
different when the saving propensity factor ($\lambda$) 
is drawn from a uniform and random distribution in [0,1] as introduced in a model proposed
by Chatterjee, Chakrabarti and Manna \cite{AKG:saving1}. Randomness in $\lambda$ is assumed to be 
quenched ({\em i.e.}, remains unaltered in time). Agents are indeed heterogeneous.
They are likely to have different (characteristic) saving propensities. 
The pair of wealth exchange equations are now written as: 
\begin{equation}\label{akg:eqn:ransave}
w_i(t+1)=\lambda_i w_i(t)+\epsilon[(1-\lambda_i)w_i(t)+(1-\lambda_j)w_j(t))], 
\end{equation}
\begin{equation*}
w_j(t+1)=\lambda_j w_j(t)+(1-\epsilon)[(1-\lambda_i)w_i(t)+(1-\lambda_j)w_j(t))]. 
\end{equation*}

\begin{figure}[htb]
\includegraphics[width=.4\textwidth]{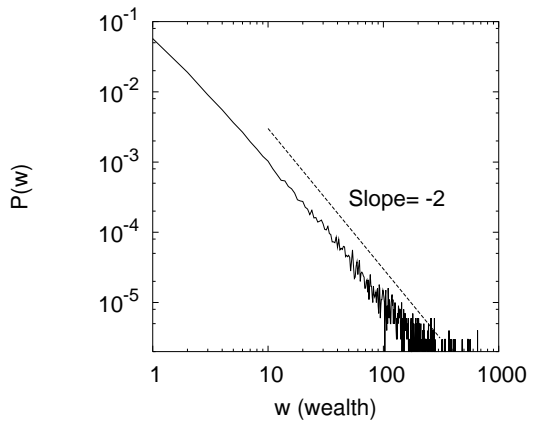}
\caption{Wealth distribution for the model of random saving propensity plotted in log-log 
scale.
A straight line with slope = -2 is drawn to demonstrate that the power law exponent is 
$\alpha=2$.
\label{akg:fig:dist_ranlam}}
\end{figure}

\noindent
A power law with 
exponent $\alpha =2$ (Pareto index $\nu=1$) is observed at the right end of the wealth 
distribution for several decades. Such a distribution is 
plotted in Fig.~\ref{akg:fig:dist_ranlam} where a straight line is drawn in the 
log-log plot with slope = -2 to illustrate the power law and the exponent. Extensive 
numerical results with different distributions in the saving propensity 
parameter $\lambda$ are reported in \cite{AKG:chak3}. Power law (with exponent $\alpha = 2$) is found to be robust.
The value of Pareto index obtained here ($\nu =1$), however, 
differs from what is generally extracted (1.5 or above) from most of the empirical data of 
income distributions (see discussions and analysis on real data by various authors in 
\cite{AKG:eco}). The present model is not able to resolve this discrepancy and it is not
expected at the outset either. 
introducing random waiting time in the interactions of agents in order to have a 
justification for a larger value of the exponent $\nu$ \cite{AKG:richmond}.  

\begin{figure}[htb]
\includegraphics[width=.4\textwidth]{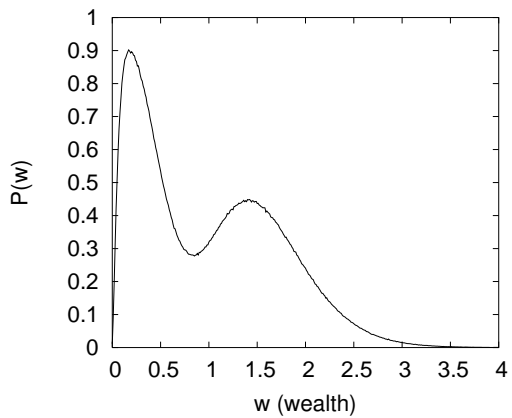}
\caption{Bimodal distribution of wealth ($w$) with fixed values of saving propensities, 
$\lambda_1$=0.2 and $\lambda_2$=0.8. Emergence of two economic classes are apparent.
\label{akg:fig:twopeak}}
\end{figure}

The distributed saving gives rise to an additional interesting feature when a special case 
is considered where the saving parameter $\lambda$ is assumed to have only two fixed values, 
$\lambda_1$ and $\lambda_2$ (preferably widely separated). A bimodal distribution 
in individual wealth results in \cite{AKG:akg1}. This can be seen from the 
Fig.~\ref{akg:fig:twopeak}. 
The system evolves towards a robust and distinct two-peak distribution as the 
difference in $\lambda_1$ and $\lambda_2$ is increased systematically. Later it is seen that 
one still gets a two-peak distribution even when $\lambda_1$ and $\lambda_2$ 
are drawn from narrow distributions centered around two widely separated values (one large and one small). 
Two economic classes seem to persist
until the distributions in $\lambda_1$ and $\lambda_2$ have got sufficient widths. 
A population can be imagined to have two distinctly different kinds of people: some 
of them tend to save a very large fraction (fixed) of their wealth and the others tend to
save a relatively small fraction (fixed) of their wealth. 
Bimodal distributions (and a polymodal distribution, in general) are, in fact, reported with real data 
for the income distributions in Argentina \cite{AKG:polymodal}. The distributions were derived at 
a time of political crisis and thus they may not be regarded as truly equilibrium distributions 
though. However, it remains an interesting possibility out of a simple model of wealth exchange. 

\subsection{Power law from mean field analysis}\label{akg:sec:meanfield}

One can have an estimate of ensemble averaged value of wealth \cite{AKG:akg3} using one of 
the above equations (\ref{akg:eqn:ransave}) in section \ref{akg:sec:ranlam}. Emergence of a power law in the 
wealth distribution can be established through a simple consideration as follows. Taking ensemble 
average of all the terms on both sides of the first eqn.~(\ref{akg:eqn:ransave}), one may write:
\begin{equation}\label{akg:eqn:meanfield-1}
\langle w_i\rangle =\lambda_i \langle w_i\rangle +\langle\epsilon\rangle[(1-\lambda_i)\langle w_i\rangle+
\langle{1\over N}\sum_{j=1}^N(1-\lambda_j)w_j\rangle] 
\end{equation}

\noindent
The last term on the right hand side is replaced by the average over agents
where it is assumed that any agent 
(here the $i$-th agent), on an average, interacts with all other agents of the system, 
allowing sufficient time to interact. This is basically a {\it mean field approach}.   
If $\epsilon$ is assumed to be distributed randomly and uniformly between 0 and 1 
then $\langle\epsilon\rangle = {1\over 2}$. 
Wealth of each individual keeps on changing due to interactions (or wealth exchange processes that 
take place in a society). No matter what the personal wealth one begins with, the time 
evolution of wealth of an individual agent at the steady state is independent of that 
initial value. This means the distribution of wealth of a single agent over time is stationary. 
Therefore, the time averaged value of wealth of any agent remains unchanged whatever the 
amount of wealth one starts with. In course of time, an agent interacts with all other agents 
(presumably repeatedly) given sufficient time. One can thus think of a number of ensembles (configurations) 
and focus attention on a particular tagged agent who eventually tends to 
interact with all other agents in different ensembles. Thus the time averaged value of 
wealth is equal to the ensemble averaged value in the steady state. 

Now if one writes 
\begin{equation}\label{akg:eqn:meanfield-2}
\langle\overline {(1-\lambda)w}\rangle \equiv \langle{1\over N}\sum_{j=1}^N(1-\lambda_j)w_j\rangle, 
\end{equation}
the above equation (\ref{akg:eqn:meanfield-1}) reduces to: 
\begin{equation}\label{akg:eqn:meanfield-3}
(1-\lambda_i)\langle w_i\rangle = \langle\overline{(1-\lambda)w}\rangle), 
\end{equation}

\noindent
The right hand side of the above equation is independent of any agent-index and the left hand side is referred 
to any arbitrarily chosen agent $i$. Thus, it can be argued that the above relation can be true for any agent 
(for any value of the index $i$) and so it can be equated to a constant. Let us now 
recognize $C = \langle\overline {(1-\lambda)w}\rangle$, a constant which is found by 
averaging over all the agents in the system and which is further 
averaged over ensembles. Therefore, one arrives at a unique relation for this model:
\begin{equation}\label{akg:eqn:meanfield-4}
w = {C\over (1 - \lambda)},  
\end{equation}

\noindent
where one can get rid of the index $i$ and may write $\langle w_i\rangle = w$ for brevity. 
The above relation is also verified numerically which is obtained by many authors in their 
numerical simulations and scaling of data \cite{AKG:chak3,AKG:gamma-support}.
One can now derive $dw = {w^2\over C}d\lambda$ from the above 
relation (\ref{akg:eqn:meanfield-4}). An agent with a 
(characteristic) saving propensity factor ($\lambda$) ends up with wealth ($w$) 
such that one can in general relate the distributions of the two: 
\begin{equation}
P(w)dw = g(\lambda)d\lambda. 
\end{equation}

\noindent
If now the distribution in $\lambda$ is considered to be uniform then 
$g(\lambda)$ = constant. Therefore, the distribution in $w$ is bound to be of the form:  
\begin{equation}
p(w) \propto {1\over w^2}.  
\end{equation}

\noindent
This may be regarded as Pareto's law with index $\alpha = 2$ which is already numerically 
verified for this present model. The same result is also obtained recently in
\cite{AKG:mohanty} where the treatment is argued to be exact. 

\subsection{Power law from reduced situation}\label{akg:sec:reduce}

From numerical investigations, it seems that the stochasticity parameter 
$\epsilon$ is irrelevant as long as the saving propensity 
parameter $\lambda$ is made random. It has been tested that the model is still able to 
produce power law (with the same exponent, $\alpha=2$) for any fixed value of $\epsilon$. 
As an example, the case for $\epsilon = 1$ is considered. The pair of wealth exchange equations (ref{akg:eqn:ransave})
now reduce to the following:
\begin{equation}\label{akg:eqn:ransave_reduce}
w_i(t+1) = w_i(t)+(1-\lambda_j)w_j(t) = w_i(t)+\eta_j w_j(t),
\end{equation}
\begin{equation*}
w_j(t+1) = w_j(t)-(1-\lambda_j)w_j(t) = (1-\eta_j)w_j(t).
\end{equation*}

\begin{figure}[htb]
\includegraphics[width=.4\textwidth]{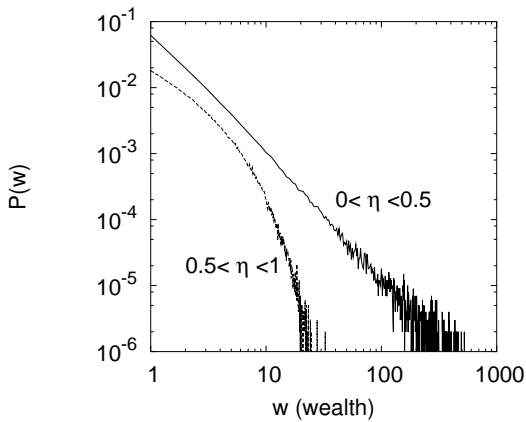}
\caption{Wealth distributions (plotted in the log-log scale) for two cases of 
the `reduced situation': (i) $0< \eta < 0.5$ and (ii) $0.5< \eta <1$ plotted in log-log scale. 
In one case, the distribution follows a power law (with exponent around $\alpha=2$) and in the other case,  
it is seen to be clearly deviating from a power law.\label{akg:fig:dist_eta}}
\end{figure}

\noindent 
The exchange amount, $\Delta w = (1-\lambda_j)w_j(t)=\eta_jw_j(t)$ is now regulated by the
parameter $\eta = (1-\lambda)$ only. If $\lambda$ is drawn from a uniform and 
random distribution in [0, 1], then $\eta$ is also uniformly and randomly distributed in 
[0, 1]. 
{\it To achieve a power law in the wealth 
distribution it seems essential that randomness in $\eta$ has to be quenched}. For `annealed' 
type disorder ({\em i.e.}, when the distribution in $\eta$ varies with time) the power 
law gets washed away (which is observed through numerical simulations). 
It has also been observed that power law can be obtained when $\eta$ is uniformly 
distributed between 0 and some value less than or equal to 1. 
As an example, $\eta$ is taken in the range between 0 and 0.5, a power 
law is obtained with the exponent around $\alpha=2$. However, when $\eta$ is taken in the 
range $0.5 < \eta < 1$, the distribution clearly deviates from power law which is evident 
from the log-log plot in Fig.~\ref{akg:fig:dist_eta}.
{\it Thus there seems to be a crossover from power law to some distribution 
with exponentially decaying tail as one tunes the range in the quenched parameter $\eta$}.

At this point, {\it two important criteria may be identified for achieving power law} within 
this reduced situation:

\begin{itemize}
\item 
The disorder in the controlling parameter $\eta$ has to be 
quenched (fixed set of $\eta$'s for a configuration of agents), 

\item 
It is required that $\eta$, when
drawn from a uniform distribution, the lower bound of that should be 0.  
\end{itemize}

\noindent
The above criteria may appear ad hoc, nevertheless have been checked by extensive numerical 
investigations. It is further checked that the power law exponent does not depend on the 
width of the distribution in $\eta$ (as long as it is between 0 and something less than 1). 
This claim is substantiated by taking various ranges of $\eta$ in which it is uniformly 
distributed. Systematic investigations are made for the cases where $\eta$ is drawn 
in [0, 0.2], [0, 0.4], $\ldots$ ,[0, 1]. 
Power laws result in in all the cases with the exponent around $\alpha = 2$. 

\section{Understanding through transition matrix}\label{akg:sec:matrix}

The evolution of wealth in the kind of two-agent wealth exchange process can be described 
through the following $2\times 2$ transition matrix ($T$) \cite{AKG:akg1}:

\[\left(\begin{array}{c}
w_i^{\prime} \\
w_j^{\prime}
\end{array}\right)=T\left(\begin{array}{c}
w_i \\
w_j
\end{array}\right),\]

\noindent 
where it is written, $w_i^{\prime}\equiv w_i(t+1)$ and $w_i\equiv w_i(t)$ and so 
on. The transition matrix ($T$) corresponding to {\it pure gambling} process (in 
section \ref{akg:sec:pure}) can be written as:

\[T=\left(\begin{array}{cc}
\epsilon & \epsilon\\
1-\epsilon & 1-\epsilon
\end{array}\right).\]

\noindent
In this case the above matrix is {\it singular} (determinant, $|T|=0$) which means the 
inverse of this matrix does not exit. This in turn indicates that an evolution through such 
transition matrices is bound to be {\it irreversible}. This property is connected to the 
emergence of exponential (Boltzmann-Gibbs) wealth distribution. The same may be perceived 
in a different way too. When a product of such matrices (for successive interactions) are 
taken, the left most matrix (of the product) itself returns: 

\[\left(\begin{array}{cc}
\epsilon & \epsilon\\
1-\epsilon & 1-\epsilon
\end{array}\right)\left(\begin{array}{cc}
\epsilon_1 & \epsilon_1\\
1-\epsilon_1 & 1-\epsilon_1
\end{array}\right)=\left(\begin{array}{cc}
\epsilon & \epsilon\\
1-\epsilon & 1-\epsilon
\end{array}\right).\]

\noindent
The above signifies the fact that during the repeated interactions of the same two agents 
(via this kind of transition matrices), the last of the interactions is what matters 
(the last matrix of the product survives) [$T^{(n)}.T^{(n-1)}\ldots T^{(2)}.T^{(1)}=T^{(n)}$]. 
This `loss of memory' (random history 
of collisions in case of molecules) may be attributed here to the path to irreversibility 
in time. 

The singularity can be avoided if one considers the following general form:

\[T_1=\left(\begin{array}{cc}
\epsilon_1 & \epsilon_2\\
1-\epsilon_1 & 1-\epsilon_2
\end{array}\right),\]

\noindent 
where $\epsilon_1$ and $\epsilon_2$ are two different random numbers 
drawn uniformly from [0, 1] (This ensures the transition matrix to be nonsingular.). 
The significance of this general form can be seen through the wealth exchange equations in the
following way: the $\epsilon_1$ fraction of wealth of the 1st agent ($i$) added 
with $\epsilon_2$ fraction of wealth of the 2nd agent ($j$) is retained by the 1st agent 
after the trade. The rest of their total wealth is shared by the 2nd agent. This may happen
in a number of ways which can be related to the detail considerations of a model. 
The general matrix $T_1$ is nonsingular as long as $\epsilon_1\neq\epsilon_2$ and then 
the two-agent interaction process remains reversible in time. Therefore, 
it is  expected to have a steady state equilibrium distribution of wealth which may deviate from 
exponential distribution (as in the case with pure gambling model). 
When one considers $\epsilon_1 = \epsilon_2$, one again gets back the pure exponential 
distribution. 
A trivial case is obtained for $\epsilon_1=1$ and $\epsilon_2=0$. The transition matrix 
then reduces to the identity matrix $I=\left(\begin{array}{cc} 1 & 0\\ 0 & 1 \end{array}\right)$
which trivially corresponds to no interaction and no evolution.

It may be emphasized that any transition matrix 
$\left(\begin{array}{cc}
t_{11} & t_{12}\\
t_{21} & t_{22}
\end{array}\right)$,
for such conserved models is bound to be of the form such that the sum of two elements of 
either of the columns has to be {\it unity by design}: $t_{11}+t_{21}=1$, $t_{12}+t_{22}=1$. 
It is important to note that whatever extra parameter, no matter, one incorporates within the framework of 
the conserved model, the transition matrix has to retain this property. 

\begin{figure}[htb]
\includegraphics[width=.4\textwidth]{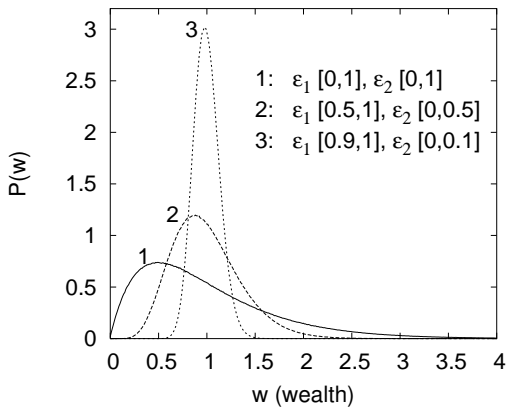}
\caption{Three normalized wealth distributions are shown corresponding to the general 
matrix $T_2$ (in general with $\epsilon_1\neq\epsilon_2$) as discussed in the text.  
Curves are marked by numbers (1, 2 and 3) and the ranges of $\epsilon_1$ and $\epsilon_2$
are indicated within which they are drawn uniformly and randomly.\label{akg:fig:dist_gen_e1_e2}
}
\end{figure}

In Fig.~\ref{akg:fig:dist_gen_e1_e2} three distributions 
(with $\epsilon_1 \neq \epsilon_2$) are plotted where
$\epsilon_1$ and $\epsilon_2$ are drawn randomly from uniform distributions with different
ranges. It is demonstrated that qualitatively different distributions are possible as the
parameter ranges are tuned appropriately.

Now let us compare the above situation with the model of equal saving propensity as discussed 
in section \ref{akg:sec:fixlam}.
With the incorporation of saving propensity factor $\lambda$, the transition matrix 
now looks like: 

\[\left(\begin{array}{cc}
\lambda+\epsilon(1-\lambda) & \epsilon(1-\lambda)\\
(1-\epsilon)(1-\lambda) & \lambda+(1-\epsilon)(1-\lambda)
\end{array}\right).\]

\noindent
The matrix elements can now be rescaled by assuming 
$\tilde\epsilon_1=\lambda+\epsilon(1-\lambda)$ and $\tilde\epsilon_2=\epsilon(1-\lambda)$ in 
the above matrix. Therefore, the above transition matrix reduces to

\[T_2=\left(\begin{array}{cc}
\tilde\epsilon_1 & \tilde\epsilon_2\\
1-\tilde\epsilon_1 & 1-\tilde\epsilon_2
\end{array}\right).\]

\noindent
Thus the matrix $T_2$ is of the same form as $T_1$. The 
distributions due to above two matrices of the same general form can now be compared 
if one can correctly identify the ranges of the rescaled elements.
In the model of uniform saving: $\lambda< \tilde\epsilon_1 <1$ and 
$0< \tilde\epsilon_2 <(1-\lambda)$ as the stochasticity parameter $\epsilon$ is 
is drawn from a uniform and random distribution in [0, 1].
As long as $\tilde\epsilon_1$ and $\tilde\epsilon_2$ are different, 
the determinant of the matrix is nonzero ($|T_2|=\tilde\epsilon_1-\tilde\epsilon_2=\lambda$). 
Therefore, the incorporation of the saving propensity factor $\lambda$ brings {\it two effects}:
\begin{itemize}
\item 
The transition matrix becomes nonsingular, 
\item 
The matrix elements $t_{11}$ (= $\tilde\epsilon_1$) and $t_{12}$ (= $\tilde\epsilon_2$)
 are now drawn from truncated domains (somewhere in [0, 1]).
\end{itemize}
Hence it is clear from the above discussion that the wealth distribution with uniform saving 
is likely to be qualitatively no different from what can be achieved with 
general transition matrices having different elements, $\epsilon_1\neq\epsilon_2$.
The distributions obtained with different $\lambda$ may correspond to that with appropriately 
chosen $\epsilon_1$ and $\epsilon_2$ in $T_1$.  

In the next stage, when the saving propensity factor $\lambda$ is distributed as in 
section \ref{akg:sec:ranlam}, the transition 
matrix between any two agents having different $\lambda$'s (say, $\lambda_1$ and $\lambda_2$) 
now looks like: 

\[\left(\begin{array}{cc}
\lambda_1+\epsilon(1-\lambda_1) & \epsilon(1-\lambda_2)\\
(1-\epsilon)(1-\lambda_1) & \lambda_2+(1-\epsilon)(1-\lambda_2)
\end{array}\right).\]

\noindent
Again the elements of the above matrix can be rescaled by 
putting $\tilde\epsilon_1^{\prime}=\lambda_1+\epsilon(1-\lambda_1)$ and 
$\tilde\epsilon_2^{\prime}=\epsilon(1-\lambda_2)$. Hence the transition matrix 
can again be reduced to the same form as that of $T_1$ or $T_2$:

\[T_3=\left(\begin{array}{cc}
\tilde\epsilon_1^{\prime} & \tilde\epsilon_2^{\prime}\\
1-\tilde\epsilon_1^{\prime} & 1-\tilde\epsilon_2^{\prime}
\end{array}\right).\]

\noindent
The determinant here is 
$|T_3|=\tilde\epsilon_1^{\prime}-\tilde\epsilon_2^{\prime}=\lambda_1(1-\epsilon)+
\epsilon\lambda_2$. 
Here also the determinant is ensured to be nonzero as all the parameters 
$\epsilon$, $\lambda_1$ 
and $\lambda_2$ are drawn from the same positive domain: [0, 1]. This means that each transition matrix 
for two-agent wealth exchange remains nonsingular which ensures the interaction process 
to be reversible in time.
Therefore, it is expected that {\it qualitatively different distributions are possible 
when one appropriately tunes the two independent elements in the general 
form of transition matrix} ($T_1$ or $T_2$ or $T_3$). 
However, the emergence of power law tail ({\it Pareto's law}) in the distribution 
can not be explained by this. Later it is examined that to obtain a power law in the 
framework of present models, it is essential that the distribution in $\lambda$ has to be 
quenched (frozen in time) which means the matrix elements in the general form of any 
transition matrix have to be quenched. In the section \ref{akg:sec:reduce}, it has 
been shown that the model of distributed saving (section \ref{akg:sec:ranlam}) is 
equivalent to a reduced situation where one needs only one variable $\eta$. 
The this case the corresponding transition matrix looks even simpler:

\[T_4=\left(\begin{array}{cc}
1 & \eta\\
0 & 1-\eta
\end{array}\right),\]

\noindent
where a nonzero determinant ($|T_4|=1-\eta\neq 0$) is ensured among other things. 

\subsection{Distributions from generic situation}\label{akg:sec:generic}

From all the previous discussions, it is clear that the 
the transition matrix (for zero sum wealth exchange) is bound to be of the 
following general form: 

\[\left(\begin{array}{cc}
\epsilon_1 & \epsilon_2\\
1-\epsilon_1 & 1-\epsilon_2
\end{array}\right).\]

\noindent
The matrix elements, $\epsilon_1$ and $\epsilon_2$ can be appropriately associated with
the relevant parameters in a model. A generic situation arrives where one can generate all
sorts of distributions by controlling $\epsilon_1$ and $\epsilon_2$. 

As long as $\epsilon_1\neq\epsilon_2$, the
matrix remains nonsingular and one achieves Gamma type distributions.
In a special case, when $\epsilon_1=\epsilon_2$, the transition matrix becomes singular and 
Boltzmann-Gibbs type exponential distribution results in.
It has been numerically checked that 
a power law with exponent $\alpha=2$ is obtained with the 
general matrix when the elements $\epsilon_1$ and $\epsilon_2$ are of the same set of 
quenched random numbers drawn 
uniformly in [0, 1]. The matrix corresponding to the reduced situation in the section \ref{akg:sec:reduce}, as discussed, 
is just a special 
case with $\epsilon_1=1$ and $\epsilon_2=\eta$, drawn from a uniform and (quenched) random 
distribution.
Incorporation of any parameter in an actual model (saving propensity, for example) 
results in the adjustment or truncation of the full domain [0, 1] from which the 
element $\epsilon_1$ or $\epsilon_2$ is drawn. 
Incorporating distributed $\lambda$'s in section \ref{akg:sec:ranlam} is equivalent to
considering the following domains: $\lambda_1< \epsilon_1 <1$ and 
$0< \epsilon_2 <(1-\lambda_2)$. 

\begin{figure}[htb]
\includegraphics[width=.4\textwidth]{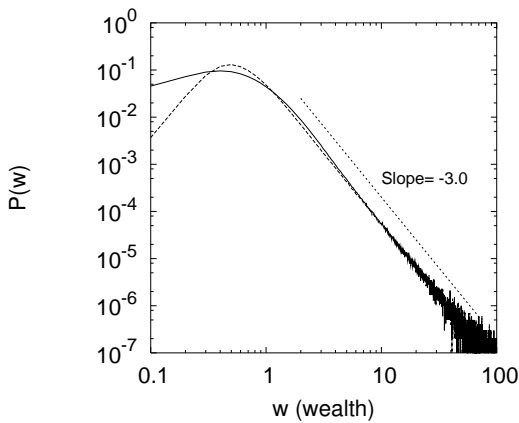}
\caption{Distribution of individual wealth ($w$) for the most general case with 
random and quenched $\epsilon_1$ and $\epsilon_2$: The elements are drawn from two separate
distributions where 
$0< \epsilon_1 <1$ and $0< \epsilon_2 <1$ in one case and in the other 
case, they are chosen from the ranges, $0.5< \epsilon_1 <1$ and $0< \epsilon_2 <0.5$. 
Both show power laws with the same exponent around 3.0 (the two distributions almost 
superpose). A straight line (with slope -3.0) is drawn to demonstrate the power law in 
the log-log scale. \label{akg:fig:dist_gen_frozen_e1_e2}} 
\end{figure}

A more general situation arrives
when the matrix elements $\epsilon_1$ and $\epsilon_2$ are of two sets of random numbers drawn separately 
(one may identify them as $\epsilon_1^{(1)}$ and $\epsilon_2^{(2)}$ to distinguish) from two uniform 
and random distributions in the domain: [0, 1].
In this case a power law is obtained with the exponent $\alpha = 3$ which is, however, distinctly
different from that obtained in `distributed saving model' in section 
\ref{akg:sec:ranlam}. 
To test the robustness of the power law, the distributions in the matrix elements are taken in 
the following truncated ranges: $0.5< \epsilon_1 <1$ and $0< \epsilon_2 <0.5$ (widths are 
narrowed down). A power law is still obtained with the same exponent ($\alpha$ close to 3). 
These results are plotted in Fig.~\ref{akg:fig:dist_gen_frozen_e1_e2}. 

It is possible to achieve distributions other than power laws as one draws 
the matrix elements, $\epsilon_1$ and $\epsilon_2$ from different domains within the range 
between 0 and 1. There is indeed a {\it crossover from power law to Gamma like distributions} 
as one tunes the elements.
It appears from extensive numerical simulations that power law disappears when both 
the parameters are drawn from some ranges that do not include the lower limit 0. For example, 
when it is considered,  $0.8< \epsilon_1 < 1.0$ and $0.2< \epsilon_2 <0.4$,  the wealth 
distribution does not follow a power law. In contrast, when $\epsilon_1$ and $\epsilon_2$ are 
drawn from the 
ranges, $0.8< \epsilon_1 < 1.0$ and $0< \epsilon_2 <0.1$, the power law distribution is back again.

It now appears that {\it to achieve a power law 
in such a generic situation, the following criteria are to be fulfilled}:

\begin{itemize}
\item 
It is essential to have the randomness or disorder in the 
elements $\epsilon_1$ and $\epsilon_2$ to be quenched,
\item 
In the most general case,
$\epsilon_1$ should be drawn from a uniform distribution whose upper bound has to be 1 
and for $\epsilon_2$ the lower bound has to be 0. Then a power law with higher exponent $\alpha =3$ is achieved. To have a power law with exponent $\alpha=2$, the matrix elements are to be 
drawn from the same distribution.
(These choices automatically make the transition matrices to be nonsingular.)
\end{itemize}

\noindent
The above points are not supported analytically at this stage.
However, the observation seems to bear important implications in terms of generation of power 
law distributions. 

When the disorder or randomness in the elements $\epsilon_1$ and $\epsilon_2$ change with 
time ({\em i.e.}, not quenched) unlike the situation just discussed above, the problem
is perhaps similar to the mass diffusion and aggregation model
by Majumdar, Krishnamurthy and Barma \cite{AKG:mass}.
The mass model is defined on a one dimensional lattice with periodic boundary condition. 
A fraction of mass from a randomly chosen site is assumed to be continually 
transported to any of its neighbouring sites at random. The total mass between the two sites 
then is unchanged (one site gains mass and the other loses the same amount) and thus the 
total mass of the system remains conserved. 
The mass of each site evolves as
\begin{equation}
m_i(t+1)=(1-\eta_i)m_i(t)+\eta_jm_j(t).
\end{equation}

\noindent
Here it is assumed that $\eta_i$ fraction of mass $m_i$ is dissociated from that site $i$ and 
joins either of its neighbouring sites $j=i\pm 1$. Thus $(1-\eta_i)$ fraction of mass $m_i$
remains at that site whereas a fraction $\eta_j$ of mass $m_j$ from the neighbouring site joins
the mass at site $i$. Now if we identify $\epsilon_1 = (1-\eta_i)$ and 
$\epsilon_2 = \eta_j$ then this model is just the same as described by the general transition
matrix as discussed so far. If $\eta_i$'s are drawn from a random and uniform 
distribution in [0, 1] then a mean field calculation (which turns out to be exact in the 
thermodynamic limit), as shown in \cite{AKG:mass}, brings out the 
stationary mass distribution $P(m)$ to be a Gamma distribution: 
\begin{equation}\label{akg:eqn:massdist}
P(m) = {4m\over {\overline m}^2}e^{-2m/{\overline m}},
\end{equation}

\noindent
where ${\overline m}$ is the average mass of the system. It has been numerically checked that 
there seems to be no appreciable change in the distribution even when the lattice is not 
considered. 
Lattice seems to play no significant role in 
the case of kinetic theory like wealth distribution models as well. 
Incidentally, this distribution with ${\overline m} = 1$ is exactly the same as the Gamma 
distribution [eqn.~(\ref{akg:eqn:gamma-1})], mentioned in section 
\ref{akg:sec:fixlam} when one considers $n = 2$ in that.
The index $n$ equals to 2 when one puts $\lambda = {1\over 4}$
in the relation (\ref{akg:eqn:gamma-2}). 

\begin{figure}[h]
\includegraphics[width=.4\textwidth]{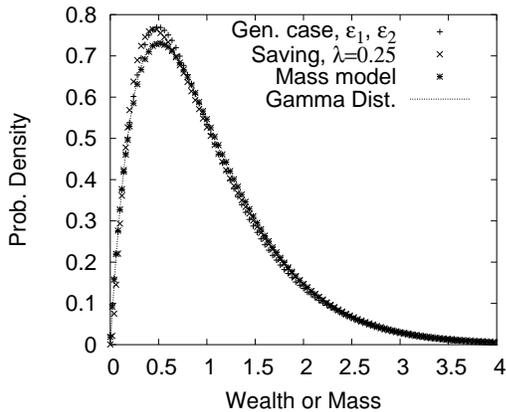}
\caption{Normalized probability distribution functions obtained for three different 
cases: (i) Wealth distribution with random and uniform $\epsilon_1$ and $\epsilon_2$ 
in [0, 1], (ii) Wealth distribution with uniform and fixed saving propensity, 
$\lambda = {1\over 4}$, (iii) Mass distribution for the model \cite{AKG:mass} in one dimensional 
lattice (as discussed in text). The theoretical Gamma distribution 
[the eqn.~(\ref{akg:eqn:massdist})] is also plotted (line draw) to have a comparison.
\label{akg:fig:gamma}}
\end{figure}

In the general situation ($\epsilon_1\neq\epsilon_2$), when both the parameters are  
drawn from a random and uniform distribution in [0, 1], the emerging distribution 
very nearly follows the above expression (\ref{akg:eqn:massdist}). Only when the 
randomness in them is quenched (fixed in time), there is a possibility of 
getting a power law as it is already mentioned.
The Gamma distribution [eqn.~(\ref{akg:eqn:massdist})] and the numerically 
obtained distributions for different cases (as discussed in the text) are plotted in 
Fig.~\ref{akg:fig:gamma} in support of the above discussions. 

\section{Role of selective interaction}\label{akg:sec:select}

\begin{figure}[htb]
\includegraphics[width=.4\textwidth]{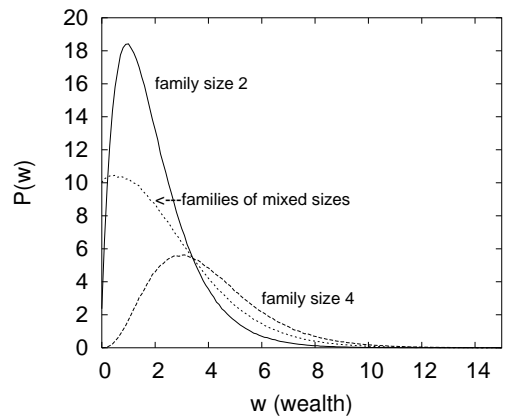}
\caption{Family wealth distributions: two curves are for families consisting of all 
equal sizes of 2 and 4. One curve is for a system of families consisting of various 
sizes between 1 and 4. The distributions are not normalized.
\label{akg:fig:family} 
}
\end{figure}

So far the models of wealth exchange processes have been considered where a pair of agents is
selected randomly. However, interactions or trade among agents in a society are often guided by 
personal choice or some social norms or some other reasons. Agents may like to interact 
selectively and it would be interesting to see how the Individual wealth distribution is 
influenced by selection \cite{AKG:akg2}. The concept of selective 
interaction is already there when one considers the formation of a family. The members of a 
same family are unlikely to trade (or interact) among each other. It may be worth to examine 
the role played by the concept of `family' in wealth distributions of families:
`family wealth distribution' for brevity. A family in a society usually consists of more than 
one agent. In computer simulation, the agents belonging to a same family are coloured to keep 
track of. To find wealth distributions of families, the contributions of the same family 
members are added up.
In Fig.~\ref{akg:fig:family} family wealth distributions are plotted 
for three cases: (i) families consist of 2 members each, (ii) 
families consist of 4 members each and (iii) families of mixed sizes between 1 and 4. 
The distributions are clearly not pure exponential, but modified exponential distributions 
(Gamma type distributions) with different peaks and different widths. This is quite 
expected as the probability of zero income of a family is zero. Modified exponential 
distribution for family wealth is also supported by fitting real data \cite{AKG:yako2}. 

\begin{figure}[htb]
\includegraphics[width=.4\textwidth]{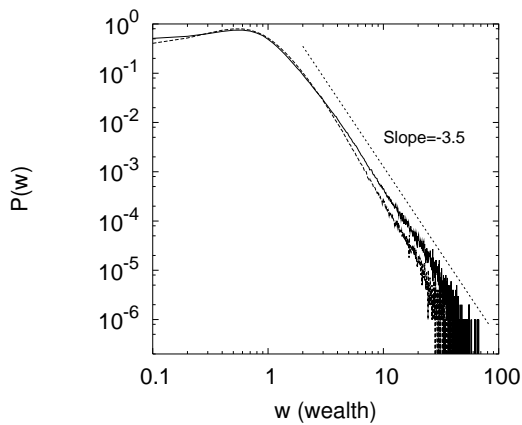}
\caption{Distribution of individual wealth with selective interaction. 
Power law is evident in the log-log plot where a straight line is drawn with
slope = -3.5 for comparison.\label{akg:fig:dist_select}}
\end{figure}

Some special way of incorporating selective interaction is seen to have a drastic effect in the individual 
wealth distribution. 
To implement the idea of `selection', a `class' of an agent is defined through an index $\epsilon$. 
The class may be understood in terms of some sort of efficiency of 
accumulating wealth or some other closely related property. Therefore, $\epsilon$'s are 
assumed to be quenched. It is assumed that during the interactions, 
the agents may convert an appropriate amount of wealth proportional to their efficiency 
factor in their favour or against. Now, the model can be understood in terms of the general 
form of equations: 
\begin{equation}\label{akg:eqn:select}
w_i(t+1) = \epsilon_i w_i(t) + \epsilon_j w_j(t), 
\end{equation}
\begin{equation*}
w_j(t+1) = (1-\epsilon_i)w_i(t) + (1-\epsilon_j)w_j(t),
\end{equation*}

\noindent 
where $\epsilon_i$'s are quenched random numbers between 0 and 1 (randomly 
assigned to the agents at the beginning).
Now the agents are supposed to make a choice to whom not to trade with. 
This option, in fact, is not unnatural in the context of a real society where individual or 
group opinions are important. There has been a lot of works on the process and dynamics of 
opinion formations \cite{AKG:stauffer,AKG:levy-solomon} in model social systems. In the present 
model it may be imagined that the `choice' is simply guided by the 
relative class index of the two agents. It is assumed that an interaction takes place when 
the ratio of two class factors remain within certain upper limit. The requirement for 
interaction (trade) to happen is then $1 < \epsilon_i/\epsilon_j <\tau$, where 
$\epsilon_i > \epsilon_j$. 
Wealth distributions for various values of $\tau$ are numerically investigated. Power laws in 
the tails of the distributions are obtained in all cases. In Fig.~\ref{akg:fig:dist_select} the 
distributions for $\tau = 2$ and $\tau =4$ are shown. 
Power laws are clearly seen with an exponent, $\alpha = 3.5$ (a straight line with slope 
around -3.5 is drawn) which means the Pareto index $\nu$ 
is close to 2.5. It is not further investigated whether the exponent ($\alpha$) actually 
differs in a significant way for different choices of $\tau$.
It has been shown that preferential behaviour \cite{AKG:prefer} 
generates power law in money distribution with some imposed conditions which allows the 
rich to get higher probability of getting richer. The rich is also favoured in a model with some 
kind of asymmetric exchange rules as proposed in \cite{AKG:sita} where a power law results in.  
The dice seems to be loaded in favour of the rich otherwise the rich can not be the rich!

\section{Measure of inequality}\label{akg:sec:ineq}

Emergence of Pareto's law signifies the existence of inequality in wealth in a population. 
Inequality or disparity in wealth or that of income is known to exist in almost all societies. 
To have a quantitative idea of inequality one generally 
plots Lorenz curve and then calculates Gini coefficient. Here the entropy approach 
\cite{AKG:kapur} is considered. The time evolution of an appropriate quantity is examined which 
may be regarded as a measure of wealth-inequality.

Let us consider $w_1,~w_2, \ldots,~w_N$ be the wealths of N agents in a system. 
Let $W = \sum_{i=1}^Nw_i$ be the total wealth of all the agents. Now $p_i=w_i/W$ can be 
considered as the fraction of wealth the $i$-th agent shares. Thus each 
of $p_i > 0$ and $\sum_{i=1}^Np_i=1$. Thus the set of $p_1,~p_2,\ldots,~p_N$ may be 
regarded as a probability distribution. The well known Shannon entropy 
is defined as the following:
\begin{equation}
S = -\sum_{i=1}^Np_i\ln p_i.
\end{equation}

\noindent
From the maximization principle of entropy it can be easily shown that the entropy ($S$) is 
maximum when
\begin{equation}
p_1=p_2=\cdots=p_N={1\over N},
\end{equation}

\noindent
giving the maximum value of $S$ to be $\ln N$ where it is a limit of 
equality (everyone possesses the same wealth). A measure of inequality should be something 
which measures a deviation from the above ideal situation. Thus one can have a measure of 
wealth-inequality to be
\begin{equation}\label{akg:eqn:inequal}
H = \ln N-S = \ln N + \sum_{i=1}^N p_i \ln p_i = \sum_{i=1}^N p_i \ln(Np_i). 
\end{equation}

\noindent
The greater the value of $H$, the greater the inequality is.

\begin{figure}[htb]
\includegraphics[width=.4\textwidth]{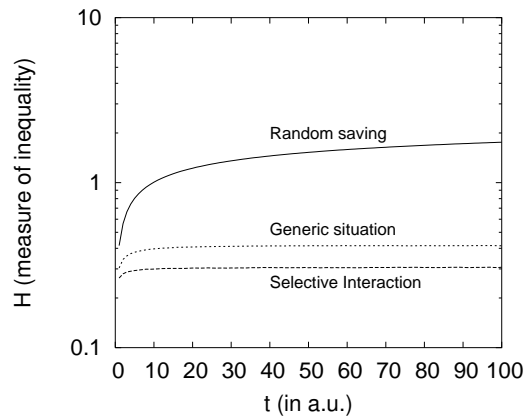}
\caption{Comparison of time evolution of the measure of inequality ($H$) in wealth for 
different models. Each `time step' ($t$) is equal to a single interaction between a pair of agents. 
Data is taken after every $10^4$ time steps to avoid clumsiness and each data point is 
obtained by averaging over $10^3$ configurations.
$Y$-axis is shown in log-scale to have a fair comparison.
\label{akg:fig:Hav_comp}}
\end{figure}

\begin{figure}[htb]
\includegraphics[width=.4\textwidth]{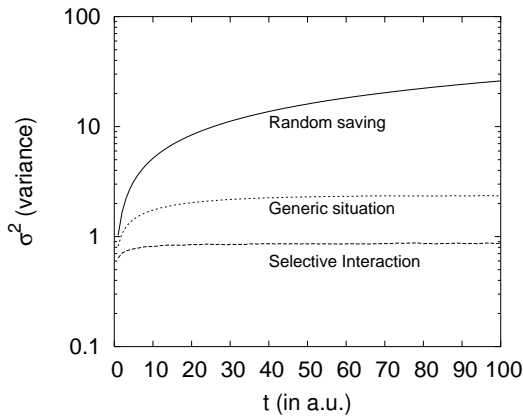}
\caption{Evolution of variance ($\sigma^2$) with `time' ($t$) for different models. 
$Y$-axis is shown in log scale to accommodate three sets of data in a same graph. 
Data is taken after every $10^4$ time steps to avoid clumsiness and each data point is 
obtained by averaging over $10^3$ configurations.
\label{akg:fig:sigav_comp}} 
\end{figure}

It is seen that the wealth exchange algorithms are so designed that the resulting 
disparity or variance (or measure of inequality), in effect, increases with time. 
Wherever power law in distribution results in, the distribution naturally 
broadens which indicates that the variance ($\sigma^2$) or the inequality measure 
[$H$ in eqn.~(\ref{akg:eqn:inequal})] should increase.
In the Fig.~\ref{akg:fig:Hav_comp} and in Fig.~\ref{akg:fig:sigav_comp} 
time evolution of inequality measure $H$ and variance $\sigma^2$ respectively 
are plotted with time for three
models to have a comparison. It is apparent that the measure of inequality in steady state 
attains different levels due to different mechanisms of wealth exchange processes, giving rise 
to different power law exponents.
The growth of variance is seen to be different for 
different models considered, which is responsible for power laws with different exponents 
as discussed in the text. The power law exponents ($\alpha$) appear to be related to the magnitudes of 
variance that are attained in equilibrium in the finite systems.

\section{Distribution by maximizing inequality}\label{akg:sec:max}

\begin{figure}[htb]
\includegraphics[width=.4\textwidth]{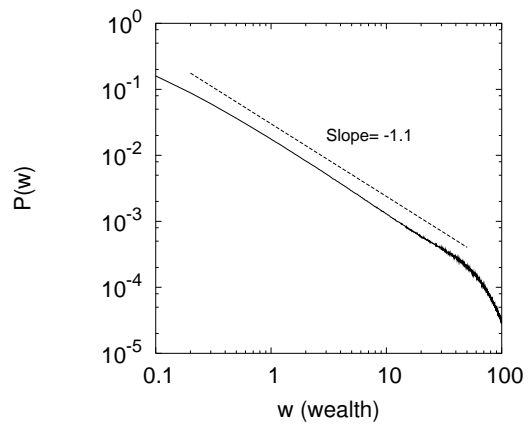}
\caption{Wealth distribution by maximizing the variance in the pure gambling model. 
Power law is clearly seen (in the log-log plot) and a straight line is drawn 
with slope = -1.1 to compare.\label{akg:fig:dist_sigmax}}
\end{figure}

It is known that probability distribution of wealth of majority is different from that of 
handful of minority (rich people). Disparity is more or less a reality in all economies. 
A wealth exchange process can be thought of within the present framework 
where the interactions among agents eventually lead to increasing 
variance. It is numerically examined \cite{AKG:akg2} whether the process of forcing the system to have ever 
increasing variance (measure of disparity) leads to a power law as it is known that power 
law is usually associated with infinite variance. 
Evolution of variance, $\sigma^2=\langle w^2\rangle-\langle w\rangle^2$ is calculated after 
each interaction in the framework of pure gambling model [the pair of equations (\ref{akg:eqn:gamble})] 
and it is then forced to increase monotonically by comparing this to the previously calculated 
value (the average value $\overline w$ is fixed by virtue of the model). 
This results in a very large variance under this imposed condition.
The inequality factor $H$ also likewise increases monotonically and attains a high value. 
A power law distribution is obtained with the exponent, $\alpha$ close to 1. 
None of the available models does bring out such a 
low value of the exponent. The variance in any of the usual models generally settles at a level 
much lower than that is obtained in such a way. 
The resulting distribution of wealth is plotted in a log-log scale 
in Fig.~\ref{akg:fig:dist_sigmax}. 

Power law, however, could not be obtained by the 
same way in the case of a non-conserved model like the following: $w_i(t+1)=w_i(t)\pm\delta$, 
where the increase or decrease ($\delta$) in wealth ($w$) of any agent is 
independent of any other. 

It has also been noted, considering some of the available models, larger the variance, 
smaller the exponent one gets. For example, the variance is seen to attain higher values 
(with time) in the model of distributed (random) saving propensities \cite{AKG:chak3} compared to 
the model of selective interaction \cite{AKG:akg2} and the resulting power law exponent $\alpha$ 
is close to 2.0 in the former case whereas it is close to 3.5 in the later. In the present 
situation the variance attains even higher values and the exponent $\alpha$ seems to be 
close to 1, the lowest among all. 

\section{Confusions and conclusions}\label{akg:sec:concl}

As it is seen, the exchange of wealth in various modes 
generates a wide variety of distributions within the framework of simple wealth exchange 
models as discussed.
In this review, some basic structures and ideas of interactions are looked at which seem to be
fundamental to bring out the desired distributions. In this kind of agent based models 
(for some general discussions, see \cite{AKG:agent}) the division of 
labour, demand and supply and the human qualities (selfish act or altruism) 
and efforts (like investments, business) which are 
essential ingredients in classical economics are not considered explicitly. 
What causes the exchange of wealth of a specific kind among agents is not important in this discussion.
Models are considered to be conserved (no inflow or outflow of money/ wealth in or from the system). It is 
not essential to look for inflation, taxation, debt, investment returns etc.
of and in an economic system at the outset for the kind of questions that are addressed here. 
The essence of complexities of interactions leading to distributions can be 
understood in terms of the simple (microscopic) exchange rules much the same way the simple logistic 
equations that went on to construct `roads to Chaos' and opened up a new horizon of thinking 
of a complex phenomenon like turbulence \cite{AKG:kadanoff}.

Some models of zero sum wealth exchange are examined here in this review. One may start 
thinking in a fresh way how the distributions emerge out of the kind of algorithmic 
exchange processes that are involved. 
The exchange processes can be understood in a general way by looking at the structure of
associated $2\times 2$ transition matrices. 
Wealths of individuals evolve to have a specific distribution in a stead state through the 
kind of interactions which are basically stochastic in nature.
The distributions shift away from Boltzmann-Gibbs like exponential to Gamma type 
distributions and in some cases distributions emerge with power law tails known as 
Pareto's law ($P(w) \propto w^{-\alpha}$). 
It is also seen that the wealth distributions seem to be influenced by personal choice. 
In a real society, people usually do not interact arbitrarily rather do 
so with purpose and thinking. Some kind of personal preference is always there which may be 
incorporated in some way or other. Power law with distinctly different 
exponent ($\alpha=3.5$, Pareto exponent $\nu=2.5$) is achieved through a certain way of 
selective interaction. The value of Pareto index $\nu$ does not correspond to what is 
generally obtained empirically. However, the motivation is not to attach much importance to 
the numerical value at the outset rather than to focus on the fact of how power 
laws emerge with distinctly different exponents governed by the simple rules of 
wealth exchange.  

The fat tailed distributions (power laws) are usually associated with large variance, which 
can be a measure of disparity. Economic disparity usually exists among a population. The detail 
mechanism leading to disparity is not always clear but it can be said to be associated 
with the emergence of power law tails in wealth distributions. 
Monotonically increasing variance (with time) can be associated with the emergence of 
power law in individual wealth distributions. 
The mean and variance of a power law distribution can be analytically derived \cite{AKG:newman} 
to see that they are finite when the power law exponent $\alpha$ is greater than 3. 
For $\alpha\le 3$, the variance diverges but then the mean is finite. In case of the models
discussed here in this review, mean is kept fixed but large or enhanced variance 
is observed in different models whenever there results in a power law. 
It remains a question of what can be the mechanisms (in the kind of discrete and 
conserved models) that generate large variance and power law tails. 
Large and increasing variance is also associated with lognormal distributions. A simple 
multiplicative stochastic process like $w(t+1)=\epsilon(t) w(t)$ can be used to explain the 
emergence of lognormal distribution and indefinite increase in variance. However, empirical 
evidence shows that the Pareto index and some other appropriate indices 
(Gibrat index, for example), generally dwindle within some range \cite{AKG:souma} indicating that 
the variance (or any other equivalent measure of inequality) does not increase forever. 
It seems to attain a saturation, given sufficient time. This is indeed the case the numerical 
results suggest. Normally there occurs simultaneous increase of variance and mean 
in statistical systems (in fact, the relationship between mean and variance goes by a 
power law as $\sigma^2 \propto {\overline w}^b$ know as Taylor's power law \cite{AKG:taylor} as 
curiously observed in many natural systems). 
In this conserved model the mean is not 
allowed to vary as it is fixed by virtue of the model. It may be the case that $\sigma^2$ 
then has to have a saturation.
The limit of $\sigma^2$ is tested through an artificial situation where the association of power law with large variance
is tested in a reverse way. 

Understanding the emergence of power law \cite{AKG:newman, AKG:reed} itself has been of great 
interest for decades. There
is usually no accepted framework which may explain the origin and wealth of varieties of 
its appearance. It is often argued that the dynamics which generate power laws is 
dominated by multiplicative processes. It is true that in an economy wealth (or money) 
of an agent multiplies and that is coupled to the large number of interacting agents. 
The generic stochastic Lotka-Volterra systems like 
$w_i(t+1)=\epsilon w_i(t)+a{\overline w(t)}-bw_i(t){\overline w(t)}$ have been 
studied \cite{AKG:levy-solomon, AKG:solomon-richmond} to achieve power law distributions in wealth.
However, these kinds of models are not discussed in this review as the basic intention 
had been to understand the ideas behind models of conserved wealth which the above is not.  

In a twist of thinking, let us imagine a distribution curve which can be stretched in any 
direction as one wishes to have, keeping the area under this to be invariant. If now the 
curve is pulled too high around the left then the right hand side is to fall off too 
quickly, exponential decay is a possible option then. 
On the other hand, if the width of it is to be stretched too far (distribution becomes fat) 
at the right hand side, it should then decay fairly slowly giving rise to a 
possible power law fall at the right end while keeping the area under the curve 
preserved. What makes such a stretching possible?
This review has been an attempt to integrate some ideas regarding models of wealth 
distributions and to reinvent things with a fresh outlook. In the way, some confusions, 
conjectures and conclusions emerged where many questions possibly have been answered with 
further questions and doubts. At the end of the day, the usefulness of this (review) may be measured 
by further curiosities and enhanced attention on the subject if at all this may generate.

\section*{Acknowledgment}
The author is grateful to Dietrich Stauffer for many important comments and criticisms at 
different stages of publications of the results that are incorporated in this review.

\end{document}